\documentstyle[twocolumn,aps,psfig]{revtex}

\begin{document}
\draft
\flushbottom
\twocolumn[
\hsize\textwidth\columnwidth\hsize\csname @twocolumnfalse\endcsname

\title{ Electron-plasmon interaction in a cylindrical mesoscopic system: important similarities with Kaluza-Klein theories}
\author{Igor I. Smolyaninov }
\address{ Department of Electrical and Computer Engineering \\
University of Maryland, College Park,\\
MD 20742}
\date{\today}
\maketitle
\tightenlines
\widetext
\advance\leftskip by 57pt
\advance\rightskip by 57pt

\begin{abstract}
Similarities between nonlinear electron-plasmon interactions in a cylindrical mesoscopic system and Kaluza-Klein theories, which stem from the analogy between the angular coordinate of a nanocylinder with the compactified coordinates of the Kaluza-Klein theories, have been considered. These similarities indicate that electron and plasmon states with non-zero angular momenta exhibit strong long-range interaction with each other via exchange of plasmons with zero angular momentum. This insight has been confirmed by finding correspondent solutions of the nonlinear Maxwell equations for interacting plasmons. Such solutions have important consequences for description of electromagnetic and transport properties of mesoscopic metallic wires, holes and rings, such as recent observation of density-dependent spin polarization in quantum wires. Numerical estimates indicate that the latter effect can be explained by the electron energy level splitting in the gyration field induced due to the magneto-optical effect by the current flowing through the quantum wire. 
\end{abstract}

]
\narrowtext

\tightenlines

1. INTRODUCTION

Electromagnetic and transport properties of mesoscopic nanowires and nanoholes are the topic of considerable current interest, due to the strong drive towards development of nanotechnology and quantum computing. While studying nanoscale science of these systems, researchers constantly meet new surprising phenomena caused by yet undiscovered quantum mechanical effects. One of such recent surprises was the discovery of anomalously high transmission of an array of nanoholes in a metal film by Ebbesen et al \cite{1}, which has been explained by light coupling to the surface plasmons on the interfaces of the film (surface plasmons are the collective excitations of conductivity electrons and the electromagnetic field). Another recent surprise was the fact that such plasmon-assisted transmission does not destroy photon entanglement \cite{2}, which makes nanohole arrays very interesting for optical quantum computing applications.  Since any quantum computer needs some interaction between the qubits, nonlinear optics of surface plasmons in nanohole arrays becomes a topic of extreme interest for many fields of science. In another recent development, strong evidences of nonlinear optical effects which indeed occur inside the nanoholes due to excitation of localized and propagating cylindrical surface plasmons have been obtained in observations of single-photon tunneling \cite{3}, and light-controlled photon tunneling effects \cite{4}. While linear optics of cylindrical surface plasmons (CSP) is relatively well understood \cite{5}, currently there is no theoretical description of their nonlinear optics. 

Another important reason for taking a close look at the nonlinear optics of nanowires and nanoholes is the current state of mesoscopic physics. There exist
numerous current contradictions between mesoscopic theories and experiment \cite{6,7,8}. At least some of these contradictions may be explained by previously unaccounted intrinsic electron-electron interactions \cite{6}. Electron-electron interaction mediated by cylindrical surface plasmons may be one of such intrinsic interactions in nanowires, since the surface plasmon electromagnetic field reaches considerable strength when the radius of the nanowire drops down to nanometer scale. As this happens, nonlinear interactions become enhanced and increasingly important (as has been observed in large number of nonlinear optical effects at nanometer-scale-rough metal surfaces, such as surface enhanced Raman scattering, etc. where enhancement factors reach up to 8-10 orders of magnitude). Unfortunately, no current mesoscopic theory takes these nonlinear interactions into account. 

The main reason for this lack of theoretical developments is the general difficulty of nonlinear systems analysis when nonlinear interactions are strong, and they can not be considered as small perturbations. In such situations, often one can only guess the general form of the solutions, based on the comparison of the nonlinear system of interest with other better understood nonlinear systems described by similar equations. Surprisingly, it appears \cite{9} that the theory of cylindrical surface plasmon (CSP) nonlinear interactions with electrons and with each other has a lot in common with a class of high energy physics theories called the Kaluza-Klein theories, which introduce extra N-4 compact and small spatial dimensions (with characteristic size on the order of the Planck length) with the symmetries of the internal space chosen to be the gauge symmetries of some gauge theory \cite{10}. The compactified extra dimensions are needed in order to formulate a unified theory, which would contain gravity together with the other observed fields. 

Similarity between the nonlinear optics of CSPs and the Kaluza-Klein theories stems from the way in which the electric charges are introduced in the original five-dimensional Kaluza-Klein theory (see for example \cite{11,12}). In this theory the electric charges are introduced as chiral (nonzero angular momentum) modes of a massless quantum field, which is quantized over the cyclic compactified fifth dimension. Electromagnetic forces between the electric charges appear as nonlinear coupling of these chiral modes, so four-dimensional electrodynamics described by the Maxwell equations may be understood as nonlinear optics of these modes. Similar Kaluza-Klein theories may be formulated in the lower-dimensional space-times. Such theories reproduce electrodynamics of electric charges in the worlds with less than three spatial dimensions. Since surface plasmons may be considered as if they live in a curved three-dimensional space-time defined by the metal interface, nonlinear optics of cylindrical surface plasmons looks as if it occurs in a space-time which besides an extended z-coordinate has a small "compactified" angular $\phi $-dimension along the circumference of the cylinder. Thus, the theory of CSP mode propagation and interaction (when the interaction is strong) may be expected to be similar to the three-dimensional Kaluza-Klein theory, and similarity between the behavior of CSP chiral modes with nonzero angular momenta and the behavior of electric charges in lower-dimensional worlds may be expected. This similarity is very useful in understanding of nonlinear optics of surface plasmons, since electromagnetic interactions of electric charges are very well understood. 

In this paper I explore the above mentioned Kaluza-Klein analogy to gain more understanding of the nonlinear mesoscopic physics of thin cylindrical metal wires and nanoholes in metal films. Using the insights gained from the Kaluza-Klein theories I am going to obtain solutions of the nonlinear Maxwell equations for interacting cylindrical surface plasmons, which behave as interacting effective charges of non-electric origin. According to these solutions, higher $(n>0)$ CSP modes posses quantized effective chiral charges proportional to their angular momenta $n$. In a metal nanowire these slow moving effective charges exhibit long-range interaction via exchange of fast massless CSPs with zero angular momentum. These zero angular momentum CSPs may be considered as massless quanta of the gyration field (the field of the gyration vector $\vec{g}$), which relates the $\vec{D}$ and $\vec{E}$ fields in an optically active medium \cite{13}. I describe the conditions for such physical picture to be valid, namely, the medium in or around the nanohole or nanowire should be chiral or optically active, and exhibit magneto-optical effect. This condition is fulfilled automatically in the case of any metal \cite{13}, which is optically active in the presence of magnetic field. This mode-coupling theory may be used in description of nonlinear optics of cylindrical nanowires and nanochannels, for example, in single-photon tunneling effect where under some circumstances there is a direct analogy between the role of Coulomb blockade in single-electron tunneling effect and the role of long-range chiral interaction of CSPs in single-photon tunneling. I will further show that the electron states with nonzero angular momenta exhibit similar long range chiral interaction via exchange of zero angular momentum CSPs. This result is of great relevance for low-temperature mesoscopic conductance of thin cylindrical wires and carbon nanotubes, which will be shown by considering the recent observation of density-dependent spin polarization in quantum wires. Thus, the nonlinear electron-electron, electron-plasmon, and plasmon-plasmon interactions may be considered in a unified way via introduction of effective chiral charges and fields, similar to the way in which unified field theories are constructed in high energy physics. It is also interesting to note that the nanohole geometries in this Kaluza-Klein description have many similarities with the space-time wormholes, which appear in many contemporary field theories. Thus, optical experiments performed on nonlinear nanoholes may become the proving ground for experimental testing of many novel ideas in theoretical physics, such as compactified extra dimensions and wormholes. Finally, I will show that so developed effective Kaluza-Klein theory of electron-plasmon interaction has strong similarities with the nonlinear electrodynamics of a rotating nonlinear-optical medium.   

2. BASIC FEATURES OF KALUZA-KLEIN THEORIES

As a first step, let us recall the basic features of the Kaluza-Klein theories.
In the original form of the theory a five-dimensional space-time was introduced where the four dimensions $x^1, ..., x^4$ were identified with the observed space-time. The associated 10 components of the metric tensor $g_{\alpha \beta }$ were used to describe gravity. After a compactified fifth dimension $x^5$ with a small circumference $L$ was added, the extra four metric components $g_{\alpha 5}$ connecting $x^5$ to $x^1, ..., x^4$ gave four extra degrees of freedom, which were interpreted as the electromagnetic potential (here we use the following convention for greek and latin indices: $\alpha = 1, ..., 4$; $i = 1, ..., 5$). An additional scalar field $g_{55}$ or dilaton may be either set to a constant, or allowed to vary. 

When a quantum field $\psi $ coupled to this metric via an equation

\begin{equation} 
\Box_5 \psi +a\psi =0
\end{equation}

is considered, where $\Box_5$ is the covariant five-dimensional d'Alembert 
operator, the solutions for the field $\psi $ must be periodic in the $x^5$ 
coordinate. This leads to the appearance of an infinite "tower" of solutions 
with quantized $x^5$-component of the momentum: 

\begin{equation}
q^5_n=2\pi n/L 
\end{equation}

where $n$ is an integer. In our four-dimensional space-time on a large scale such solutions with $n\neq 0$ interact with the electromagnetic potential $g_{\alpha 5}$ as charged particles with an electric charge $e_n$ and mass $m_n$:

\begin{equation}
e_n = \hbar q_n(16\pi G)^{1/2}/c
\end{equation}

\begin{equation}
m_n = \hbar (q_n^2-a)^{1/2}/c
\end{equation}

where $G$ is the gravitational constant (see for example the derivation in 
\cite{11}). For the purposes of discussion below let us follow this derivation 
when an angular coordinate $\phi ^5$ varying within an interval from 0 to 
$2\pi $ is introduced, so that 

\begin{equation}
\phi ^5 = 2\pi \frac{x^5}{L}   
\end{equation}

Now the metric can be written as

\begin{equation}
ds^2=g_{\alpha \beta }dx^{\alpha }dx^{\beta }+2g_{\alpha 5}dx^{\alpha }d\phi 
^5+g_{55}d\phi ^5d\phi ^5,
\end{equation} 

Here we are not interested in possible spatial dependence of $g_{55}$ and 
consider all $g_{i5}$ components to be independent of $\phi ^5$. 
Equation (1) with $a=0$ for the quantum field $\psi $ in this metric should be 
written as

\begin{eqnarray}
\frac{\partial }{\partial x^{\alpha }}(g^{\alpha \beta }\frac{\partial \psi 
}{\partial x^{\beta }})+\frac{\partial }{\partial x^{\alpha }}(g^{\alpha 
5}\frac{\partial \psi }{\partial \phi ^5})+ \nonumber \\
\frac{\partial }{\partial \phi ^5}(g^{5\alpha }\frac{\partial \psi }{\partial 
x^{\alpha }})+\frac{\partial }{\partial \phi ^5}(g^{55}\frac{\partial \psi 
}{\partial \phi ^5})=0
\end{eqnarray}

Since we assume that the $g_{i5}$ do not depend explicitly on $\phi ^5$, we 
should not address the terms like $\partial g^{i5}/\partial \phi ^5$. Thus, we 
may search for the solutions in the usual form as
$\psi = \Psi (x^{\alpha })e^{iq\phi ^5}$, where periodicity in $\phi ^5$ 
requires $q_n=n$. As a result, we obtain

\begin{equation}
\Box \psi -q_n^2\frac{1-g_{\alpha 5}g^{\alpha 5}}{g_{55}}\psi+2iq_ng^{\alpha 
5}\frac{\partial \psi}{\partial x^{\alpha }}+iq_n \frac{\partial g^{\alpha 
5}}{\partial x^{\alpha }}\psi = 0
\end{equation} 

This is the same as the Klein-Gordon equation in the presence of an 
electromagnetic field: in four-dimensional space-time it describes a particle of 
mass

\begin{equation}
m=\frac{2\pi \hbar q_n}{cg_{55}^{1/2}(q_n)},
\end{equation}

which interacts with a vector field $g^{\alpha 5}$ through a quantized charge 
$e_n \sim q_n$. In this theory the conservation of the electric charge is a 
simple consequence of the conservation of the $x^5$-component of the momentum. 
Similar Kaluza-Klein theory may be formulated in a three-dimensional space-time, 
which has one compactified spatial dimension. Again, quantized $\phi $-component 
of the momentum will play a role of an effective charge, which interacts with a 
two-component vector field $g^{\alpha 3}$.  

It is well-known that Maxwell equations in a general curved space-time 
background $g_{ik}(x,t)$ are equivalent to the macroscopic Maxwell equations in 
the presence of matter background with some nontrivial electric and magnetic 
permeability tensors $\epsilon _{ik}(x,t)$ and $\mu _{ik}(x,t)$ \cite{14}. Thus, 
strong similarity between the results of the three-dimensional Kaluza-Klein 
theory and the solutions of Maxwell equations in some quasi-three-dimensional 
cylindrical waveguide geometries may be expected: in both cases we consider a 
massless field in a geometry which has a compactified cyclic $\phi $-dimension 
in the presence of nontrivial space-time curvature or permeability tensors, 
respectively. This similarity is expected to be especially strong in the case of 
surface plasmon waveguides, since surface plasmons live in (almost) three-
dimensional space-times on the metal interfaces.  

3. NONLINEAR OPTICS OF CYLINDRICAL SURFACE PLASMONS

In this chapter the similarity described above will be confirmed by finding the respective solutions of the nonlinear Maxwell equations for cylindrical surface plasmons. 
Let us first consider general properties of Maxwell equations in a medium (media), which has cylindrically symmetric geometry. The insight gained from the previous discussion tells us that in order to look similar to the three-dimensional Kaluza-Klein theory, the medium should discriminate between left- and right- circular polarized waves (the waves which have opposite angular 
momenta, and thus expected to posses opposite effective Kaluza-Klein charges). 
This means that the medium should be chiral or optically active. There are 
different ways of introducing optical activity (gyration) tensor in the 
macroscopic Maxwell equations. It can be introduced in a symmetric form, which 
is sometimes called Condon relations \cite{15}:

\begin{equation}  
\vec{D}=\epsilon \vec{E}+\gamma\frac{\partial \vec{B}}{\partial t}
\end{equation}

\begin{equation}  
\vec{H}=\mu ^{-1}\vec{B}+\gamma\frac{\partial \vec{E}}{\partial t}
\end{equation}

Or it can be introduced only in an equation for $\vec{D}$ (see \cite{13,16}). In 
our consideration I will follow Landau and Lifshitz \cite{13}, and for simplicity use only the following equation valid in isotropic or cubic-symmetry materials:

\begin{equation}  
\vec{D}=\epsilon \vec{E}+i\vec{E}\times \vec{g} ,
\end{equation}

where $\vec{g}$ is called the gyration vector. If the medium exhibits magneto-
optical effect and does not exhibit natural optical activity $\vec{g}$ is 
proportional to the magnetic field $\vec{H}$:

\begin{equation}  
\vec{g}=f\vec{H} ,
\end{equation}

where the constant $f$ may be either positive or negative. For metals in the 
Drude model at $\omega >>eH/mc$

\begin{equation}  
f(\omega )= -\frac{4\pi Ne^3}{cm^2\omega ^3}=-\frac{e\omega _p^2}{mc\omega ^3} ,
\end{equation}

where $\omega _p$ is the plasma frequency and $m$ is the electron mass \cite{13}.

Initially, let us consider a medium with $\vec{g}=\vec{g}(r,z,t)$ directed along 
the $\phi $-coordinate. Such a distribution may be produced, for example, in a 
medium exhibiting magneto-optical effect around a cylindrical nanowire if a 
current is passed through the wire. After simple calculations we obtain a wave 
equation in the form:

\begin{equation}  
\vec{\nabla }\times \vec{\nabla }\times \vec{B}=-\Delta \vec{B}=-\frac{\epsilon 
}{c^2}\frac{\partial ^2\vec{B}}{\partial t^2}+\frac{i}{c}\frac{\partial 
(\vec{\nabla }\times [\vec{E}\times\vec{g}])}{\partial t}  
\end{equation}

The z-component of this wave equation for a solution $\sim e^{in\phi }$ is:

\begin{eqnarray}
\frac{1}{r}\frac{\partial }{\partial r}(r\frac{\partial B_z}{\partial 
r})+\frac{\partial ^2B_z}{\partial z^2}-\frac{\epsilon \partial 
^2B_z}{c^2\partial t^2}-\frac{n^2}{r^2}B_z+ \nonumber \\
i\frac{ng}{rc}\frac {\partial (iE_z)}{\partial t}+\frac{in}{cr}(\frac{\partial 
g}{\partial t})(iE_z)=0
\end{eqnarray}
 
It can be re-written in the form similar to equations (1) and (8) as follows:

\begin{equation}
\hat{a_r} B_z+\Box _2B_z-\frac{n^2}{g_{\phi \phi 
}}B_z+in(\frac{g}{r})(\frac{\partial iE_z}{c\partial t})+in\frac{\partial 
(g/r)}{c\partial t}E_z=0,
\end{equation}

where $\hat{a_r}$ plays the role of factor $a$ in equation (1), and $g_{\phi \phi }=r^2$. Similarity of equations (8) and (17) becomes more evident if we 
identify the Kaluza-Klein vector field $g^{0\phi }$ as $g/r$, disregard terms 
higher than linear in $g$, and recall that for cylindrical surface plasmon modes 
$iE_z = \alpha B_z$, where the coefficient of proportionality $\alpha $ is real 
and is determined by the boundary conditions \cite{5}.
The missing factor of 2 in front of the fourth term in equation (17) originates 
from the asymmetric way of introducing the gyration in the macroscopic Maxwell 
equations (see equations (10-12) and the relevant discussion). Thus, from the 
macroscopic Maxwell equations we arrive to a picture of $n>0$ waveguide modes 
interacting with the "gyration potential" $\sim g/r$ via quantized effective "chiral charges" proportional to the angular momenta of the modes. 

The dispersion law and electromagnetic field distribution of cylindrical surface 
plasmon modes of a cylindrical metal wire may be found in \cite{5}. The $\omega 
(k)$ of the $n=0$ CSP mode goes down to $\omega =0$ approaching the light line 
$\omega =kc/\epsilon ^{1/2}$ from the right, as $k\rightarrow 0$. The field of 
this mode has only the following nonzero components: $E_r$, $E_z$, and $H_{\phi 
}$, thus satisfying our initial requirements for the direction of $\vec{g}$, and 
if we recall equation (13), we notice that in the media exhibiting magneto-
optical effect, such as the cylindrical metal wire itself, the higher $n>0$ CSP 
modes interact with the $H_{\phi }$ field of the $n=0$ CSPs via their effective 
"chiral charges" proportional to $n$. Thus, CSP quanta with $n=0$ may be 
considered as massless (in the $k\rightarrow 0$ limit) quanta of the "gyration 
field". The dispersion laws of the CSP modes with $n>0$ start at some nonzero 
frequencies and intersect the light line (become nonradiative and de-couple from 
free-space photons) at some finite $\omega $. The modes of different $n$ are 
well separated from each other, and there is no crossing. In the $k\rightarrow 
\infty $ limit the $\omega (k)$ of all the CSP modes saturates at $\omega 
=\omega _p/2^{1/2}$. Thus, the group velocity $d\omega /dk$ of the higher modes 
goes to 0 in this limit: the "charged" quanta are slow.  

Let us now derive an analog of the Poisson equation for the "gyration potential" 
and "chiral charges" around the cylindrical metal wire. Let us search for the 
solutions of the nonlinear wave equation (15) in the form $\vec{B}=\vec{B_0}+\vec{B_n}$ and $\vec{E}=\vec{E_0}+\vec{E_n}$, where $\vec{B_0}$ and $\vec{B_n}$ are the CSP 
fields with zero and nonzero ($n$) angular momenta, respectively, and the 
"gyration field" in the magneto-active media is obtained in a self-consistent 
manner as $\vec{g}=f(\vec{H}+\vec{B_0}+\vec{B_n})$, where $\vec{H}$ is a 
constant external field. We are interested in the solution for the field 
$\vec{B_0}$ in the limit $\omega _0 \rightarrow 0$ in the presence of the 
$\vec{B_n}$ field, so that in the found solution the $\vec{B_n}$ field will act as a source of $\vec{B_0}$. The resulting nonlinear Maxwell equation may be simplified assuming that the field $\vec{B_n}$ is supposed to be the solution of linear Maxwell equation, and the terms proportional to $f^2$ and higher may be 
neglected. As a result, we obtain:
\begin{eqnarray}  
\Delta \vec{B_0}=-\frac{if}{c}\frac{\partial (\vec{\nabla }\times 
[\vec{E_n}\times\vec{B_n}])}{\partial t}-\frac{if}{c}\frac{\partial (\vec{\nabla 
}\times [\vec{E_0}\times\vec{B_n}])}{\partial t}- \nonumber \\
\frac{if}{c}\frac{\partial (\vec{\nabla }\times 
[\vec{E_n}\times\vec{B_0}])}{\partial t}      
\end{eqnarray}
Since the fields $\vec{B_0}$ and $\vec{B_n}$ are not supposed to be coherent, their products disappear after time averaging, and we are left with the effective Poisson equation for the nonlinear interaction of CSPs:

\begin{equation}  
\Delta \vec{B_0}=\frac{f\omega _n}{c}\vec{\nabla }\times 
[\vec{E_n}\times\vec{B_n}]=\frac{4\pi f\omega _n}{c^2}\vec{\nabla }\times 
\vec{S_n} ,
\end{equation}

where $\vec{S_n}$ is the Pointing vector of the CSP field with the nonzero ($n$) 
angular momentum. As an interesting consequence of this equation, let us note 
that in an isotropic medium 

\begin{equation}  
\vec{\nabla}\times (\vec{\nabla }\times \vec{B_0}-\frac{4\pi f\omega _n 
}{c^2}\vec{S_n})=0 
\end{equation}

This observation lets us to conjecture that electric currents may also be the sources of the chiral field, and that the electron states with nonzero angular momenta may act as chiral charges. We will discuss this extension of the CSP nonlinear optics later in this paper.  

Using the same approximations as before, we can also derive an analog of the 
Gauss theorem for the "chiral charges". Let us consider a cylindrical volume $V$ 
around a cylindrical metal wire (see Fig.1), such that the side wall of the 
volume $V$ is located very far from the wire and the CSP fields are zero at this 
wall. 

\begin{figure}[tbp]
\centerline{
\psfig{figure=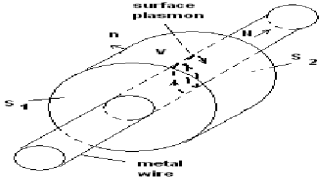,width=8.5cm,height=5.0cm,clip=}}
\caption{ Schematic view of a cylindrical metal wire, which supports cylindrical 
surface plasmon propagation. $\vec{N}$ is the chosen direction of the wire.
}
\label{fig1}
\end{figure}

Using the vector calculus theorem we can write

\begin{equation}
\int_{V}\vec{\nabla }\times \vec{S}d^3x=\int_{S}\vec{n}\times 
\vec{S}da=\int_{S_2}\vec{N}\times \vec{S}da-\int_{S_1}\vec{N}\times \vec{S}da,
\end{equation}

where $S$ is the closed two-dimensional cylindrical surface bounding $V$, with 
area element $da$ and unit outward normal $\vec{n}$ at $da$, $S_1$ and $S_2$ are 
the front and the back surfaces of $V$, and $\vec{N}$ is the chosen direction of 
the wire. Using equation (20) we obtain

\begin{eqnarray}
\int_{V}\frac{4\pi f\omega _n}{c^2}\vec{\nabla }\times 
\vec{S}d^3x=\int_{S_2}\vec{N}\times [\vec{\nabla }\times \vec{B_0}]da- \nonumber 
\\
\int_{S_1}\vec{N}\times [\vec{\nabla }\times \vec{B_0}]da
\end{eqnarray}

Since $\vec{N}\times [\vec{\nabla }\times \vec{B_0}]=\partial B_{0\phi }/\partial z$, we see that a "chiral charge" produces a local step in the 
"gyration field". If this effective Gauss theorem would be precise at any 
distance, we would come up with an unphysical result that a localized "chiral 
charge" would produce an infinite field $B_{\phi }$ at infinity. In reality, one 
has to take into account all the effects that we neglected while deriving the 
effective Poisson equation and Gauss theorem, such as higher than quadratic 
terms in the nonlinear Maxwell equations, and also take into account final 
lifetime of the cylindrical surface plasmons. Nevertheless, the underlying 
Kaluza-Klein mechanism of the "chiral charge" interaction lends credibility to 
the conclusion that CSPs of a cylindrical metal nanowire exhibit very strong 
long-range interaction. The case of cylindrical nanochannel is somewhat 
different, since in this case the zero angular momentum mode has nonzero 
frequency in the limit $k\rightarrow 0$ (in other words it acquires some 
effective mass), so that interaction of higher mode quanta via exchange of zero-
angular momentum CSPs becomes more short-ranged. Similar mode-coupling theory 
may be further developed to describe mode propagation and interaction in other 
chiral optical waveguides.   

Let us estimate the strength of chiral charges interaction. Using equations (17) 
and (22) within the approximations described above the potential energy of two 
equal chiral charges $n$ separated by distance $z$ can be written as

\begin{equation}
W=z\frac{2\pi \hbar \omega _nf^2nS_{n\phi }}{\epsilon rc},
\end{equation}

where $S_{n\phi }=n\omega _n(\epsilon E_z^2+B_z^2)/(4\pi \beta _n^2r)$ is the 
$\phi $-component of the Pointing vector, 
$\beta _n^2=\epsilon \omega _n^2/c^2-k_n^2$ is defined by the dispersion law of the $n$th mode, and $r$ is the radius 
of the cylindrical metal wire. Assuming $f(\omega )$ from (14) and $\beta _n\sim k_n$ the order of magnitude estimate of $W$ may be written as

\begin{equation}
W=zn^2(\frac{e^2}{\hbar c})\hbar \omega _n (\frac{\omega _p^4}{\omega 
_n^4})(\frac {\lambda }{r})\frac{\hbar ^2}{r^3m^2c^2},
\end{equation}

where $\lambda $ is the wavelength of $\omega _n$ light in vacuum. As may be 
expected, the nonlinear optical interaction of CSP modes grows inversely 
proportional to $r^4$, and achieves considerable strength in nanometer scale 
metallic nanowires and nanochannels even without addition of nonlinear 
dielectrics inside a nanochannel or around a nanowire. If we assume $\hbar 
\omega _n=1$eV, $\hbar \omega _p=7$eV, and $r=1$nm (disregarding nonzero skin 
depth) the potential energy of two unit chiral charges is of the order of $W\sim 
2\times 10^4z$eV/cm, so even interaction of single CSP quanta separated by 
submicrometer distances is considerable. The effect of this interaction on the ground state of the nanowire will be discussed later in this paper.

A very useful physical picture of the chiral interaction may be obtained using the analogy between the Maxwell equations in a given gravitational field \cite{17}, and the Maxwell equations in a chiral medium. Relationship between the $\vec{D}$ and $\vec{E}$ fields in a gravitational field with the metric $g_{ik}$ can be written in the form

\begin{equation}  
\vec{D}=\frac{\vec{E}}{h^{1/2}}+\vec{H}\times \vec{g} ,
\end{equation}

where $h=g_{00}$ and $g_{\alpha }=-g_{0\alpha }/g_{00}$. For an electromagnetic wave this relationship is very similar to equation (12) if we identify 
$\epsilon $ as $1/h^{1/2}$, and take into account different conventions for the use of imaginary numbers used in \cite{13} and \cite{17}. Thus, the space-time itself behaves as a chiral optical medium if $\vec{g}\neq 0$. Such a space-time region can be described locally as a rotating coordinate frame with an angular velocity \cite{17}

\begin{equation}  
\vec{\Omega }=\frac{ch^{1/2}}{2}\vec{\nabla }\times \vec{g} 
\end{equation}
 
Similar vector $\vec{\Omega }$ field can be defined for any chiral medium regardless of the nature of its optical activity (natural or magnetic field induced). Thus, nonlinear electrodynamics of such medium may be understood as if there is a distribution of local angular rotation $\vec{\Omega }$ field in it. It is clear that in the presence of $\vec{\Omega }\neq 0$ (as is the case for a zero-momentum CSP propagating along the metal wire, due to the $H_{\phi }$ field of the CSP and eq.(13)) any particle with nonzero angular momentum (CSP or electron) acquires additional energy $\vec{L}\vec{\Omega }$, where $\vec{L}$ is the angular momentum of the particle. This is the physical meaning of the terms in equations (16) and (17) describing the chiral interaction of the CSP angular momenta with the gyration field. At the same time, the field of CSPs with zero angular momentum describes small spatial oscillations of the $\vec{\Omega }$ field. Thus, zero-angular-momentum CSPs may be considered as quanta of the effective angular rotation field in the optically active nonlinear medium, and, according to the solution of the nonlinear Maxwell equations described by equation (19), CSPs with nonzero angular momenta (the chiral charges) act as the sources of the $\vec{\Omega }$ field.

4. NONLINEAR ELECTRON-PLASMON INTERACTION

Let us now extend the above ideas to the description of nonlinear effects in electron-plasmon interaction. While linear Schrodinger equation is appropriate to determine the lowest order effects of the CSP field on the individual electron wave functions, such a description would not be appropriate in the case of nanowires and nanoholes, since plasmon field strength may become considerable and nonlinear optics of the materials must be taken into account. Thus, higher order nonlinear interaction terms of the nonlinear medium quantum electrodynamics must be considered. At the same time, since plasmons are collective excitations of the electron density and the electromagnetic field, collective effects must be taken into account if a proper description of electron-plasmon interaction must be obtained. Thus, we meet the general difficulty of the mesoscopic physics of finding the ground state of a one-dimensional electron system, which is described in \cite{8} and the references therein. These arguments indicate considerable difficulties in constructing nonlinear theory of electron-plasmon interaction. On the other hand, reasonable success of using the insights gained from the analogy with the Kaluza-Klein theories in finding solutions of the nonlinear Maxwell equations for the interacting CSPs encourage us to develop similar analogy for the more complicated nonlinear electron-plasmon interactions. 

The logical continuation of the ideas described above is to consider both the chiral charges of the CSPs and the normal electric charges in a symmetric and unified way, in which electron-electron, electron-plasmon, and nonlinear optical plasmon-plasmon interaction in a thin metal wire is described as if it happens in a four-dimensional space-time, which besides the extended z-coordinate has two compactified cyclic spatial dimensions: the $\phi $-dimension along the circumference of the metal cylindrical wire, and the cyclic fifth dimension of the original Kaluza-Klein theory. Similar to high energy physics where the search for the underlying universal symmetries between different particles and fields has been a very fruitful approach, we may expect that such a unified description may bring about new deep insights into the mesoscopic physics. Such insights are highly necessary taking into account numerous current contradictions between mesoscopic theories and experiment \cite{6,7,8}. At least some of these contradictions may be explained by previously unaccounted intrinsic electron-electron interactions \cite{6}. Electron-electron interaction via exchange of cylindrical surface plasmons may be one of such interactions. 

Let us now follow the general Kaluza-Klein recipe, and try and build a unified theory of electron-plasmon interaction in a thin metal wire by considering a four-dimensional space-time, which besides the extended z-coordinate has two compactified cyclic spatial dimensions: the $\phi $-dimension along the circumference of the metal cylindrical wire, and the cyclic fifth $\theta $-dimension of the original Kaluza-Klein theory. As has been seen above, the effects of the radial coordinate can be included in the factor $a$ of equation (1), so we will neglect the radial coordinate of the real physical wire for the sake of simplicity. We will also assume temperature to be very low, so that interactions of only a few electrically charged quaziparticles can be considered, while the rest of the electrons in the metal are taken into account via electric and magnetic permeability tensors $\epsilon _{ik}$ and $\mu _{ik}$ of the metal, which in turn are taken into account via the space-time metric $g_{ik}$. Thus, the effective metric can be written as

\begin{eqnarray}
ds^2 = c^2dt^2 - dz^2 - R^2d\phi ^2 - r^2d\theta ^2+ 2g_{02}cdtd\phi + 
\nonumber \\
2g_{03}cdtd\theta + 2g_{12}dzd\phi + 2g_{13}dzd\theta + 2g_{23}d\theta d\phi
\end{eqnarray} 

where $R(z)$ and $r$ are the radii of the cylindrical wire and the compactified original fifth Kaluza-Klein dimension, respectively. Here we consider all the $g_{0i}$ and $g_{1i}$ components to be independent of $\phi $ and $\theta $. 
Equation (1) with $a=0$ for a quantum massless scalar field $\psi $ in this metric should be written as 

\begin{equation}
\frac{\partial }{\partial x^i}(g^{ik}\frac{\partial \psi }{\partial x^k})=0,
\end{equation}

where $i,k=0, ... , 3$. The field $\psi $ is considered to be scalar in order to make the consideration as simple as possible (for a vector field each component of the field will need to satisfy equation (28)). We will search for the solutions in the usual form as
$\psi = \Psi (x^{\alpha })e^{iq_2\phi }e^{iq_3\theta }$, where $\alpha =0,1$, and periodicity in $\phi $ and $\theta $ requires $q_2=n_2$ and $q_3=n_3$ ($n_2$ and $n_3$ are integer). As a result, we obtain

\begin{eqnarray}
(\frac{\partial ^2}{c^2\partial t^2}-\frac{\partial ^2}{\partial z^2})\psi -(q_2^2g^{22}+q_3^2g^{33}+2q_2q_3g^{23})\psi+ \nonumber \\ 
2iq_2g^{\alpha 2}\frac{\partial \psi}{\partial x^{\alpha }}+2iq_3g^{\alpha 3}\frac{\partial \psi}{\partial x^{\alpha }}+iq_2 (\frac{\partial g^{\alpha 2}}{\partial x^{\alpha }})\psi+iq_3 (\frac{\partial g^{\alpha 3}}{\partial x^{\alpha }})\psi = 0
\end{eqnarray} 

This is a two-dimensional Klein-Gordon equation describing a two-component quantized superchiral charge $(q_2;q_3)$ in the presence of external $g^{\alpha 2}(t,z)$ and $g^{\alpha 3}(t,z)$ vector fields (the name superchiral is chosen to emphasize the unified description of electrons and plasmons). The $q_3$ component of the superchiral charge and the $g^{\alpha 3}$ field correspond to the quantized electric charge $e \sim q_3$ and the electromagnetic field, respectively, while the $q_2$ component of the charge and the $g^{\alpha 2}$ field correspond to the chiral charge and the gyration field described above. Thus, we have obtained a unified symmetric description of both types of charges and their interactions. For example, it is easy to see that the term $2q_2q_3g^{23}\psi $ in equation (29) corresponds to the orbital magnetic moment of the electric charge $q_3$, which interacts with the axial magnetic field described by $g^{23}$. At the same time, equation (29) indicates that the axial magnetic field is not the only field acting on the electric charges which posses nonzero orbital momenta. According to (29), such states respond to the gyration field $g^{\alpha 2}$ in the same way as the CSPs with nonzero angular momenta ($n>0$). In other words, a rotating electric charge has an effective chiral charge, and it exhibits long-range interactions with other rotating electric charges and $n>0$ CSPs via exchange of fast massless CSPs with zero angular momentum.     

After we have understood the general symmetry and structure of the electron-plasmon interaction in the Kaluza-Klein model, we can try and reformulate this description using normal language of nonlinear Maxwell and Schrodinger equations. Let us start from the effective Poisson equation for the chiral charges and the gyration field in the cylindrical wire geometry in the form of equation (20). By looking at the expression under the parenthesis, it is clear that the electric current $\vec{j}$ should act as an additional source of $\vec{B_0}$, so that the effective Poisson equation takes the form

\begin{equation}  
\Delta \vec{B_0}=\frac{4\pi f\omega _n}{c^2}\vec{\nabla }\times 
\vec{S_n}+\frac{4\pi }{c}\vec{\nabla }\times \vec{j} 
\end{equation}

Thus, both the chiral charges of the CSPs and the rotating electric charges enter this equation in a symmetric way, and act as the sources of the gyration field in agreement with the Kaluza-Klein analogy. We can also conjecture that the electron's spins should also enter as the sources in this equation, although the proof of this conjecture can only be obtained in a complete nonlinear medium quantum electrodynamics treatment, which is beyond the scope of this paper.

As a second step, let us consider how the field of a zero-angular-momentum CSP (gyration field) acts on a rotating electric charge. The field of this CSP mode has only the following nonzero components: $E_r$, $E_z$, and $H_{\phi }$ \cite{5}, and may be described in cylindrical coordinates $(r,\phi ,z)$ by the vector potential of the form $\vec{A}=(A_r, 0, A_z)$, so that $E_r=\partial A_r/\partial t$, $E_z=\partial A_z/\partial t$, and $H_{\phi }=\partial A_r/\partial z-\partial A_z/\partial r$. The linear Shrodinger equation with the vector potential of this form can be used in evaluation of CSPs effect on the phase of the electron wave function. Unfortunately, the chiral interaction of interest is due to nonlinear optical effects, and arises as higher order terms in QED of a nonlinear optical medium. However, the analogy with the electrodynamics of a rotating medium described above makes it physically clear that any particle with nonzero angular momentum (CSP or electron) acquires additional energy $\vec{L}\vec{\Omega }$ in the gyration field, where $\vec{L}$ is the angular momemtum of the particle. Thus, the Kaluza-Klein analogy again looks justified in predicting the chiral field interaction with rotating electric charges.

As has been shown above, in a thin metal wire the chiral interaction of the CSP chiral charges via exchange of zero-angular-momentum CSPs becomes quite noticeable at large distances. Unlike the Coulomb interaction, which is screened by the presence of other free electrons, the chiral interaction does not experience much screening at short distances, due to its general weakness: we may say that the chiral charges are adiabatically free, similar to the behavior of quarks at short distances. However, because of the one-dimensional nature of the chiral interaction, it does not depend on the distance between the chiral charges unless the CSP decay (finite free propagation length) is taken into account. The strong frequency dependence of the constant $f$ in the equation (14) may even lead to the increase of the chiral interaction with distance (since $f$ grows proportional to $\lambda ^3$, as long as $\omega >>eH/mc$).
These factors may make the chiral electron-electron, electron-plasmon and plasmon-plasmon interactions an important mechanism in mesoscopic transport phenomena, especially in long samples.     

A number of experiments to check the importance of chiral interactions may be suggested. For example, the CSP spectrum of a cylindrical metal wire may be changed by periodic modulation of the shape of the wire, and the effects of this change on different mesoscopic properties may be studied. Comparison of transport properties of wires with different cross-sections may also be performed. Breaking of the cylindrical symmetry of the wire may be described as an external chiral field, which affects the chiral electron-plasmon interaction, but does not remove it completely. An external chiral field may also be created by coating the wire with a layer of chiral optical material or external illumination with circular-polarized light. In all these cases the chiral electron energy level splitting may be studied. If the energy level splitting is high enough, one can start to talk about para- or dia-chiral response of the samples. 

The chiral interaction is especially important in possible applications of the nanohole arrays in quantum computing. A nanohole geometry can be described by equation (27) by choosing $R(z)\rightarrow \infty $ at $z=0$ and $z=d$, where $d$ is the metal film thickness. This choice of metric makes it look like a wormhole of the field theories, which connects two "flat" three-dimensional surface-plasmon space-times on the interfaces of the metal film. Since the structure of the equation (29) is not affected by z-dependence of R, the language of chiral interaction is the proper language to describe surface plasmon interaction around the nanoholes in a nanohole array. This interaction quickly fades away from the nanoholes, but it provides weak coupling of the qubits necessary for the quantum computer operation, while the external illumination of the nanohole array may act as a computer program.  These are just a few of possible experimental checks and developments of the ideas described in this paper.

I should also mention that the density-dependent spin polarization recently observed in zero magnetic field in ultra-low-disorder quantum wires \cite{8} may be explained as a manifestation of chiral energy level splitting, especially if we recall the observation of the authors that this effect is more pronounced in longer samples and at higher electron densities \cite{18} (as we have seen from the previous discussion, longer samples increase the importance of the chiral interaction, while higher electron densities increase the plasma frequency, and, hence, the value of the constant $f$ in eq.(13)). Electric current flowing through the quantum wire during measurements may create some initial gyration field $\vec{g}=f\vec{H}$ circulating around the wire, and hence, the $\vec{\Omega}\sim \vec{\nabla }\times \vec{g}$ field directed along the wire. 
This will induce the initial chiral asymmetry between the left- and right- nonzero angular momenta of electrons and plasmons (the energy level splitting of the order of $\Delta E=\hbar \Omega $). Although this initial splitting is probably too weak, it creates some predominance of chiral charges of one sign with respect to another, so one may expect some net chiral charge of the same sign separated by distances as large as the length of the sample. Taking into account our earlier estimate of the energy of the chiral charge interaction of the order of $W\sim 2\times10^4\times l$ eV/cm, where $l$ is the sample length, we see that the chiral ordering (and, hence, spin polarization) may be expected in long quantum wires. Although more precise theoretical treatment of this effect will be necessary in the future, these arguments indicate that the chiral energy level splitting is an observable effect, and it may in fact already been observed. 

In conclusion, similarities between nonlinear electron-plasmon interactions in a cylindrical mesoscopic system and Kaluza-Klein theories, which stem from the analogy between the angular coordinate of a nanocylinder with the compactified coordinates of the Kaluza-Klein theories, have been considered. These similarities indicate that electron and plasmon states with non-zero angular momenta exhibit strong long-range interaction with each other via exchange of plasmons with zero angular momentum. This insight has been confirmed by finding correspondent solutions of the nonlinear Maxwell equations for interacting plasmons. Such solutions have important consequences for description of electromagnetic and transport properties of mesoscopic metallic wires, holes and rings, such as recent observation of density-dependent spin polarization in quantum wires. Numerical estimates indicate that the latter effect can be explained by the electron energy level splitting in the gyration field induced due to the magneto-optical effect by the current flowing through the quantum wire.


\begin{references}

\bibitem{1} T.W. Ebbesen $et$ $al.$, Nature 391, 667 (1998).
\bibitem{2} E. Altewischer $et$ $al.$, Nature 418, 304 (2002).
\bibitem{3} I.I. Smolyaninov, A.V. Zayats, A. Gungor, and C.C. Davis, Phys.Rev.Lett. 88, 187402 (2002).

\bibitem{4} I.I. Smolyaninov, A.V. Zayats, A. Stanishevsky, and C.C. Davis, Phys.Rev.B 66, 205414 (2002).

\bibitem{5} U. Schroter and A. Dereux, Phys.Rev.B 64, 125420 (2001).

\bibitem{6} P. Mohanty and R.A. Webb, Phys.Rev.Letters 88, 146601 (2002).

\bibitem{7} R. Deblock $et$ $al.$, Phys.Rev.Letters 89, 206803 (2002).

\bibitem{8} D.J. Reilly $et$ $al.$ Phys.Rev.Letters 89, 246801 (2002).

\bibitem{9} I.I. Smolyaninov, cond-mat/0209662

\bibitem{10} M.J. Duff, B.E.W. Nilsson, and C.N. Pope, Phys.Rep. 130, 1 (1986).

\bibitem{11} A. Chodos and S. Detweiler, Phys.Rev. D 21, 2167 (1980).

\bibitem{12} I.I. Smolyaninov, Phys.Rev.D 65, 047503 (2002).

\bibitem{13} L.D. Landau and E.M. Lifshitz, Electrodynamics of Continuous Media 
(Pergamon, New York, 1984).

\bibitem{14} W. Schleich, M.O. Scully, in: G. Grynberg, R. Stora (Eds.), New Trends in Atomic Physics (North-Holland, Amsterdam, 1984) p.997.

\bibitem{15} D.F. Nelson, J.Appl.Phys. 86, 5348 (1999).

\bibitem{16} M. Born and K. Huang, Dynamical Theory of Crystal Lattices 
(Clarendon, Oxford, 1966), pp.336-338.

\bibitem{17} L.D. Landau and E.M. Lifshitz, Field Theory (Pergamon, New York, 1984).

\bibitem{18} D.J. Reilly $et$ $al.$, Phys.Rev.B 63, R121311 (2001). 

\end{references}
\end{document}